# Mapping Innovation Networks: A Network-Based Approach to Actor Heterogeneity in National Innovation Systems

Dawoon Jeong,[a,b] , Taewon Kang[c], Saerom Si[d] Sangnam Lee[e],* and Wonsub Eum[f, ]*

[a] *Kellogg School of Management, Northwestern University, Evanston, IL, 60208, USA*

[b] *Northwestern Institute on Complex Systems, Evanston, IL, 60208, USA*

[c] *Graduate School of National Public Policy, Chungnam National University, Dajeon, Korea*

[d] *Technology Management, Economics and Policy Program, Seoul National University, 1 Gwanak-ro, Gwanak-gu, Seoul, 08826, Republic of Korea*

[e] *Korea Institute of Science & Technology Evaluation and Planning (KISTEP), Wonjung-ro, Maengdong-myeon, Eumseong-gun, Chungcheongbuk-do, 1339, Republic of Korea*

[f] *Department of Global Business, Rikkyo University, 3 Chome-34-1 Nishiikebukuro, Toshima City, Tokyo, 171-8501, Japan*

*\* (Corresponding authors)*

**Abstract**

The Triple Helix model has provided a foundational framework for analyzing National Innovation Systems by highlighting the roles of universities, industries, and government research institutes. However, increasing heterogeneity within these actor groups limits the explanatory power of typological approaches. This study introduces a capability-based network methodology that maps the structural relationships among innovation actors based on the similarity of their research and development (R&D) capabilities. Drawing on Economic Complexity Theory, we measure each actor's revealed comparative advantage (RCA) across scientific and technological fields and construct an R&D Actor Space—a proximity-based network that reflects the relational configuration of innovation capacities. Applying this method to Korean R&D data, we uncover a stratified system in which central, highly diversified universities coexist with more specialized firms and government institutes. Network analysis reveals assortative and unequal structures, and hierarchical clustering further highlights layered subgroupings. By moving beyond categorical classification, this capability-based network approach provides a scalable and generalizable tool for analyzing structural complexity within national innovation systems.

**Keywords:** National innovation system, economic complexity, Triple R&D Actor Space

# 1. Introduction

The National Innovation System (NIS) encompasses the institutions, actors, and interactions that collectively influence the creation, diffusion, and application of economically valuable knowledge and technologies (Freeman, 1987; Lundvall, 1992; Nelson, 1993). Within this system, a diverse set of innovation actors—including universities, firms, and government research institutes—work together to shape national capacities for technological development (Lundvall, 2016; M. A. Shapiro et al., 2010; Nelson, 1993).

To describe and analyze the interactions among innovation actors, the Triple Helix (TH) model has become a widely used framework. By categorizing actors in the NIS into three primary groups—university (U), industry (I), and government (G)—the TH model has provided a foundation for interpreting the structural evolution and innovation dynamics of national systems (Etzkowitz & Leydesdorff, 1998; Leydesdorff & Etzkowitz, 1998a, 1998b; Leydesdorff & Meyer, 2006; Shapiro, 2007; Park & Leydesdorff, 2010; Yoon, 2015). However, as innovation actors within the NIS have become increasingly heterogeneous, the traditional TH model faces limitations in fully capturing the complexity of contemporary innovation systems. In response, researchers have proposed N-tuple helices models that expand the actor base beyond UIG to include non-profit organizations, international institutions, financial actors, and societal stakeholders (Leydesdorff, 2012, 2020; Arranz et al., 2020; Vetsikas & Stamboulis, 2023).

While such expansions address part of the problem, the heterogeneity within traditional actor groups—particularly within universities, industries, and government research institutes (UIG)—is often overlooked. Innovation actors increasingly diverge in their technological specialization, research capacity, and institutional roles. For example, universities differ widely in their disciplinary strengths and types of knowledge production;

firms vary by size, sector, and R&D intensity; and government research institutes may function as either mission-driven entities or as service-oriented providers for industrial support (Corsaro et al., 2012; Gulbrandsen, 2011). These internal variations imply that heterogeneity in the NIS is not only a result of adding new actor types, but also stems from the increasing differentiation of capabilities among existing actors. Consequently, relying solely on broad classifications such as UIG may obscure the fine-grained diversity of innovation dynamics within the system. What is needed is a framework that allows for more granular and relational observation of innovation actors and their capabilities—beyond categorical typologies.

This study argues that a more fundamental shift in perspective is needed: rather than further subdividing actor types or introducing new helices, we propose a *network-based lens* that enables direct observation of the relational structure among innovation actors based on their *capability proximity*. To implement this lens, we draw on *economic complexity theory* (ECT), which has proven effective for measuring the capabilities of countries, regions, and institutions across economic (Hidalgo & Hausmann, 2009), technological (Kogler et al., 2013, 2017; Rocchetta et al., 2022), knowledge (Chinazzi et al., 2019; Guevara et al., 2016), and skills (Lee et al., 2025a, 2025b ; Muneepeerakul et al., 2013; Walter et al., 2024).

We apply ECT to R&D activity data in order to construct what we term the *R&D Actor Space*: a R&D capability-based network where innovation actors are connected based on the similarity of their revealed comparative advantage (RCA) across scientific and technological fields. This network allows us to map the structural configuration of innovation actors without presupposing actor types or groupings, and to examine the degree of connectivity, inequality, and clustering that emerges from their actual innovation behavior.

By shifting the analytical focus from categorical classification to relational observation, this study contributes a new framework for examining the internal dynamics of national innovation systems. While our empirical application focuses on the Korean NIS, the methodology is generalizable to other contexts and datasets. The results reveal structural patterns—such as assortative connectivity and unequal capability clusters—that deepen our understanding of how innovation actors are organized and how they interact.

The remainder of this paper is structured as follows. Section 2 reviews the relevant literature on the models for heterogeneous agents in NIS and economic complexity theory. Section 3 presents the methodological framework for constructing the R&D Actor Space. Section 4 applies the model to Korean R&D data and analyzes the resulting network structure. Section 5 discusses the implications of the findings and concludes the study.

## 2. Literature review

### 2.1. Heterogeneous agents in the NIS

Innovation within the NIS emerges from the interactions of various institutional actors with distinct roles, resources, and capabilities. Especially, recent innovation systems have evolved into complex structures, as a wider range of social and economic actors participate and traditional actors become more diversified and segmented. While the TH model—centered on universities, industry, and government—has served as a foundational framework for conceptualizing these interactions (Etzkowitz & Leydesdorff, 2000; Leydesdorff & Etzkowitz, 1998; Leydesdorff, 2003; Leydesdorff & Sun, 2009; Y. H. Lee & Kim, 2016), recent shifts in the structure and behavior of innovation actors reveal its analytical limitations (Leydesdorff & Meyer, 2006; Arranz et al., 2020).

Interactions among universities, industries, and governments (including government research institutes) have become significantly more multifaceted and differentiated than in the past, making it difficult for the traditional TH model to fully capture these dynamics (Zadegan et al., 2025). The TH model has often treated universities, industries, and governments as homogeneous entities, which limits its ability to reflect the heterogeneity and diverse capabilities of each actor. Consequently, analyzing only the structural networks among UIG actors is insufficient to understand the innovation capabilities of the NIS (Arranz et al., 2020).

In particular, the traditional classification of innovation actors into universities, firms, and government research institutes (GRIs) masks substantial internal heterogeneity. Universities differ significantly in terms of disciplinary focus, research intensity, and systemic roles—ranging from knowledge production to entrepreneurship and regional development (Saad et al., 2015; Datta et al., 2019; Fonseca & Nieth, 2021). GRIs also exhibit institutional hybridity, functioning as either mission-oriented institutes focused on national strategic technologies or service-oriented research providers that support industrial innovation (Gulbrandsen, 2011; Giannopoulou et al., 2019).

Firms present perhaps the most diverse set of innovation behaviors. Large firms tend to dominate innovation efforts, while small and medium-sized enterprises (SMEs) often face capability constraints (Kaufmann et al., 2003). Firms also differ in their innovation orientation: some focus on product innovation, while others specialize in process innovation and actively engage in inter-organizational collaboration (Bellucci & Pennacchio, 2016). This diversity has been well documented in earlier studies on industrial innovation, which highlight firm-specific paths of technological development and their use of external knowledge (Abernathy & Utterback, 1978; Utterback & Abernathy, 1975; Ahuja & Katila, 2001; Meyer et al., 1997; Ulrich, 1995). Firms may absorb and apply technologies developed by universities and GRIs (Krishna, 2019), or integrate them into sector-specific production systems (Kolomytseva & Pavlovska, 2020).

These differences are not peripheral but central to understanding how innovation emerges and diffuses within the NIS. As Corsaro et al. (2012) emphasize, actors' roles and capabilities are becoming increasingly divergent, shaping collaboration networks and knowledge flows. Yet conventional typologies such as UIG remain too coarse to capture these structural dynamics. While some studies attempt to overcome this limitation by expanding the TH model into N-tuple models—adding non-profit, financial, or international actors (Leydesdorff, 2012, 2020; Arranz et al., 2020; Vetsikas & Stamboulis, 2023)—these extensions primarily address actor variety rather than the internal heterogeneity within existing categories.

This study shifts the focus from classification to relational observation. To establish effective innovation policies and design actor-specific strategies, it is essential to adopt a quantitative approach that captures the detailed characteristics of innovation agents and the structure of their interactions. Therefore, we propose a network-based framework—R&D Actor Space—to capture the configuration of innovation capabilities among actors. Rather than relying on predefined categories, this approach emphasizes the structural proximity of innovation actors based on the similarity of their research and development (R&D) profiles, enabling finer-grained observation of heterogeneity and inequality within the NIS.

## 2.2. Economic complexity theory

ECT is a theory that seeks to understand and explain economic development across countries using a complex system approach, where a country's level of development is expressed in terms of the complexity of the economic structure it is forming (Hidalgo et al., 2007; Hidalgo and Hausmann, 2009). ECT measures a country's overall degree of economic complexity, via revealed comparative advantage in the products that it exports (Balassa, 1965). The more countries simultaneously possess comparative advantages in two different products, the higher the commonality of actors' capabilities for exporting the two products; two products with a high commonality of capabilities are more related (Boschma, 2005; Hidalgo et al., 2007). Complex products connected to various products in terms of proximity are high value-added products, and countries with a high comparative advantage in complex products are countries with a high complexity of economic structure and a high level of development. The relationship between export products is represented as a network, with products as nodes and proximity between products as links. The network is named Product Space (Desmarchelier et al., 2018; Hidalgo et al., 2007).

ECT is a generalized theory that explains the development of actors and their activities. ECT has been applied to a variety of data beyond national export activity. Specifically, it has been applied to regions' patenting representing inventing activities and to build knowledge spaces and suggest knowledge diversification strategies (Asheim, 2019; Balland et al., 2019; Eum and Kang, 2022; Eum and Lee, 2022a; Kogler et al., 2013, 2017); it has also been applied to research articles representing research activities and to build research spaces and analyze relationships and comparative advantages between research fields (Chinazzi et al., 2019; Guevara et al., 2016).

ECT's appropriateness in analyzing actors and their complex capability relationships in various data scope has been validated by various empirical studies (Hidalgo, 2021), yet there is a room to expand the ECT from the perspective of innovation capability by linking the theory with the TH model. To this end, our study applies ECT to observe the heterogeneous innovation actors that comprise the NIS and examine their innovation activity through a complex system approach. The detailed ECT methodology and its application to NIS are described in detail in Section 3.

## 3. Research Model

### 3.1. Mapping the innovation capability of agents: Specialization Matrix

Actors' heterogeneous roles in the system can be observed through the activities they perform. Innovation actors such as universities, firms and GRIs in the NIS perform innovation activities through R&D. R&D effort can be equated with knowledge innovation, that is, the amount of R&D expenditures undertaken by actors in NIS can be viewed as the amount of innovation activities (Verba, 2022).

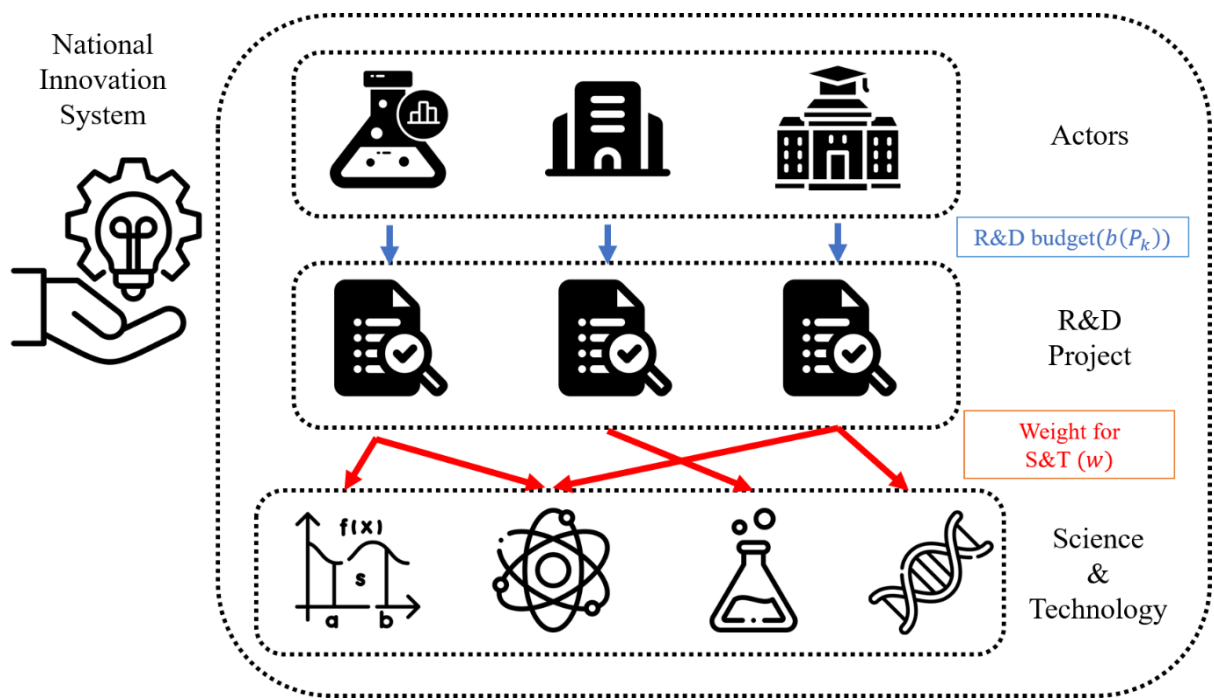

**Figure 1.** Conceptual framework of activity in R&D

As shown in Figure 1, each of the actors that comprise the NIS ($a$) conduct innovation activity through one or more R&D ($P_k$) for Science & Technology (S&T), such as math, physics, chemistry, and biology. Each R&D may belong to a single S&T, or it may belong to one or more different S&Ts. The $w$ is the weight that the R&D belongs to for each S&T. The amount of innovation activity is represented by the research expenditure on R&D project $k$ ($b(P_k)$). Specifically, $X_{as}$ is the amount of actor a's innovation activity for S&T's, expressed by Eq. 1.

$$X_{as} = \sum_{P_k} w_k^s b(P_k) \qquad \text{(Eq .1)}$$

$w_k^s$ is the weight on S&T($s$) of the R&D k ($P_k$) performed by actor a, with a value between 0 and 1. An R&D can have a weight in more than one S&T, and all weights add up to 1. Among all R&D conducted by $a$, the R&D expenditure of the projects corresponding to a specific S&T ($s$) is multiplied by the corresponding weight ($w_k^s b(P_k)$) of all R&D conducted by a in a specific S&T (s) multiplied by the corresponding weights.

$$RCA_{as} = \frac{X_{as}/\sum_{s'} X_{as'}}{\sum_{a'} X_{a's}/\sum_{a's'} X_{a's'}} \qquad \text{(Eq .2)}$$

RCA is a measure of an actor's comparative advantage in a particular activity (Balassa, 1965). RCA measures the concentration required in an activity by calculating the share of that activity in the actor's basket of activities and the overall share of that activity in the activities of all actors. An actor is considered to have a comparative advantage in an activity if its RCA value is greater than or equal to 1 (Hidalgo, 2021). RCA has been used to measure countries' comparative advantage in export (Eum and Lee, 2019, 2022a, 2022b; Hidalgo, 2021), a region's technological comparative advantage (Kogler et al., 2013) and a researcher's comparative advantage in a research area (Guevara et al., 2016). In this study, we utilize RCA to measure the comparative advantage of innovation activity in the S&T of innovation actors in the NIS.

$$M = (M_{as})$$

$$M_{as} = 1 \text{ if } RCA_{as} \geq 1 \text{ otherwise } 0$$

(Eq .3)

The S&Ts where innovation actors in the NIS have a comparative advantage (RCA≥1) can be represented by a matrix, known as the specialization matrix (*M*) (Eq. 3). *M* is a matrix with rows consisting of UIGs and columns consisting of S&Ts, where the component of the actor in row a's S&T field in column s $M_{as}$ is equal to 1 if the actor in row a has a comparative advantage ($RCA_{as} \geq 1$); otherwise, $M_{as} = 0$.

## 3.2. Heterogeneous capability of agents: Diversity & Ubiquity

Based on the *M* presented in Eq. 3, we can measure the diversity and average ubiquity of actors' innovation capabilities.

Diversity of agent $a$  $(k_{a,0}) = \sum_s M_{as}$

Ubiquity of S&T $s$  $(k_{s,0}) = \sum_a M_{as}$

(Eq .4)

Average ubiquity of S&Ts owned by agent $a$  $(k_{a,1}) = \frac{1}{k_{a,0}} \sum_s M_{as} k_{s,0}$

Eq. 4 represents the diversity and average ubiquity for NIS actors, utilizing the method of reflection (Hidalgo and Hausmann, 2009). $k_{a,0}$ is the number of S&Ts in which *a* possesses comparative advantage, which indicates the diversity of *a*'s innovation capabilities. $k_{s,0}$ is the number of actors with capabilities in S&T *s*, that is, $k_{s,0}$ is a ubiquitous S&T in which many actors have capabilities. $k_{a,1}$ is the average ubiquity of S&Ts owned by *a* with capabilities. Through $k_{a,0}$ and $k_{a,1}$ in Eq. 4, we can measure the diversity and average ubiquity of actors' innovation capabilities in the NIS.

## 3.3. Connection between agents: Proximity and R&D Actor Space

In this section, we present a methodology for building an R&D Actor Space, a model that represents the relationships between actors in the NIS as a network based on the similarity of their innovation capabilities. In ECT, a space can be constructed with trade codes (Hidalgo et al., 2007), research fields (Chinazzi et al., 2019; Guevara et al., 2016), and patent codes as nodes which represent activities (Kogler et al., 2013), or a space can be built using actors such as countries (Bahar et al., 2014; Hidalgo, 2021). In this study, each of the actors conducting R&D are set as nodes and an agent represented its innovation capability as a vector.

$$v_a = [RCA_{a1}, RCA_{a2}, RCA_{a3}, \ldots, RCA_{aN}]$$

$$\phi_{ij} = cosine\ similarity(v_i, v_j)$$

(Eq .5)

Eq. 5 provides a vector representation of the actor *a*, utilizing RCA values for the S&Ts. Then, the cosine similarity of different actor vectors is defined as actors' proximity. Previous studies have also utilized the capabilities vector of actor ($v_a$) in Eq. 5. However, we design proximity slightly differently. In previous studies, vectors are log-transformed, then correlated and used to measure proximity (Bahar et al., 2014). Previous study analyzed export activity between countries, most countries have at least a small amount of export values, so that the RCA value is not 0. Therefore, the error due to log transformation is less likely to occur. However, in the case of R&D in the NIS, there are many S&Ts in which individual actors do not conduct R&D, and the RCA value is often measured as 0. Therefore, similarity, which can be used as an alternative to correlation, is used to represent proximity in this study. Since the vectors are sparse vectors with many zero values, cosine similarity is used to measure the similarity of the RCA composition between actors.

$$R\&D\ Space = G(V, E)$$

$$V = \{a | a \in Actors\ in\ NIS\}$$

$$E = \{e_{ij} | e_{ij} = 1\ if\ cosine\ similarity(v_i, v_j) \geq threshold;\ e_{ij} = 0\ otherwise\}$$

(Eq. 6)

A network can be built by connecting individual actors as nodes to form a link if the proximity of capabilities between them is above a certain threshold; a link should not be formed otherwise (Eq. 6). The network built in this way is named R&D Actor Space. R&D Actor Space helps us to intuitively observe the innovation capabilities' relationship between actors in the NIS. In addition, we can analyze the overall structural characteristics of the NIS by performing a degree analysis on the R&D Actor Space.

Degree is the number of links a node has in a network. By observing the distribution and correlation of degrees, we can understand the network's structural characteristics. The probability distribution of degree $n$ values ($P(n)$) is proportional to $n^{-\gamma}$ (Eq. 7); $\gamma$ is called the degree exponent. Degree exponent refers to the network's degree of preferential attachment (PA). If $\gamma=0$, it is a random network with no PA degree, and if $\gamma$ is large, it is a scale-free network with a large PA property. In general, if *2<γ<3*, it is considered to have a scale-free network characteristic (A.-L. Barabási, 2013; A. L. Barabási et al., 2000; A. L. Barabási and Albert, 1999; A. L. Barabási and Bonabeau, 2003).

$$P(n) \sim n^{-\gamma}$$

$$log(P(n)) \sim (-1) \times \gamma \, log(n)$$

(Eq .7)

Correlation function ($k_{nn}(n)$) is average degree of neighbors ($N(n)$) for degree *n* nodes. Correlation function ($k_{nn}(n)$) has a distribution proportional to $n^\mu$ (Eq. 8); $\mu$ is called the correlation exponent. If $\mu > 0$, it is an assortative network in which nodes with large degrees are connected to nodes with large degrees; if $\mu = 0$, it is a neutral network. If $\mu < 0$, then it is a disassortative network in which nodes with larger degrees are connected to nodes with smaller degrees (A. L. Barabási, 2005; A.-L. Barabási, 2013). Degree exponent ($\gamma$) and Correlation exponent ($\mu$) quantitatively analyze the structural characteristics of R&D Actor Space.

$$k_{nn}(n) = \frac{1}{|N(n)|} \sum_{j \in N(n)} Degree(j)$$

$$k_{nn}(n) \sim n^\mu$$

$$log(k_{nn}(n)) \sim \mu \, log(n)$$

(Eq .8)

## 3.4. Hierarchy in R&D Actor Space

The R&D Actor Space represents a network structure in which nodes correspond to innovation actors and links reflect their proximity in innovation capabilities. In this network, actors with similar R&D profiles are more likely to be connected, forming local clusters that reflect structural similarities in innovation behavior. These proximities are derived from pairwise comparisons of revealed comparative advantages across scientific and technological fields, and offer a relational basis for observing structural patterns in the NIS.

To further examine the internal configuration of the R&D Actor Space, this study applies a clustering methodology to identify how innovation actors are hierarchically structured based on capability similarity. The purpose is not to assign actors into fixed clusters but to explore how groups emerge naturally through relational proximity. This approach enables the identification of nested subgroupings that are otherwise obscured by predefined classifications.

Among various clustering techniques—such as K-means, DBSCAN (Ester et al., 1996), and hierarchical clustering—we adopt hierarchical clustering due to the nature of our data. Methods like K-means and DBSCAN assume that observations exist in a coordinate-based metric space, which is incompatible with our pairwise similarity-based framework. Hierarchical clustering, in contrast, requires only a proximity matrix and constructs a tree-like structure that captures multi-level relationships between actors.

The clustering procedure follows a bottom-up, agglomerative approach. Initially, the two most similar actors are merged into a cluster. Subsequent merges occur iteratively based on proximity until all actors are connected. The resulting dendrogram provides a visual summary of these merging steps, highlighting how tightly connected actors form local cores, and how more peripheral actors are progressively incorporated. This visual representation of nested structures allows us to interpret degrees of similarity and separation across the network.

Rather than seeking a single optimal number of clusters, our objective is to utilize hierarchical clustering as a structural lens to assess the underlying stratification of innovation actors in the NIS. The dendrogram enables us to examine the extent to which actors form centralized cores or remain isolated, offering insight into the degree of heterogeneity and potential core-periphery dynamics. This approach is particularly valuable in contexts where categorical distinctions (e.g., Triple Helix types) are insufficient to capture relational complexity among innovation actors.

## 4. Empirical Results

### 4.1. Data: Korean R&D project

In this section, empirical research is conducted to verify the appropriateness of the methodology proposed in section 3. The data for the empirical research are R&D data from the National Science and Technology Information System (NTIS) of South Korea. South Korea is a representative country that has grown from a developing to developed country through active R&D investment. Therefore, R&D data from NTIS is actively utilized in NIS and R&D empirical studies (Jeong et al., 2018; S. Lee et al., 2022; Yang et al., 2014). In this study, we analyze 63,305 R&D projects in Korea undertaken in 2021. The organizations which principal investigators subject to are defined as the actors of the R&D. The innovation actors are classified into four types in the NTIS: university, firm (Industry), GRI (Government) and others. Of the 593 innovation actors, 45 have more than one overlapping type. For example, the Korea Advanced Institute of Science and Technology is categorized as both a university and GRI.

| Number of R&D | | | Investment of R&D | | |
|---|---|---|---|---|---|
| Category | Name | Number of R&D projects | Category | Name | Investment of R&D projects (MKW) |
| University | Seoul National University | 2,319 | Firms | Korea Aerospace Industries Co. | 886,414 |
| University | Yonsei University | 1,868 | GRI | Electronics and Telecommunications Research Institute | 580,592 |
| University | Korea University | 1,600 | University | Seoul National University | 447,427 |
| University/GRI | Korea Advanced Institute of Science and Technology | 1,496 | GRI | Korea Aerospace Research Institute | 305,613 |
| University | Sungkyunkwan University | 1,186 | University | Yonsei University | 298,820 |
| University | Kyungpook National University | 1,038 | University/GRI | Korea Advanced Institute of Science and Technology | 291,781 |
| University | Pusan National University | 993 | University | Korea University | 286,674 |
| University | Hanyang University | 992 | GRI | Korea Advanced Institute of Science and Technology | 259,982 |
| GRI | National Academy of Agricultural Sciences | 809 | GRI | Institute for Basic Science | 256,963 |
| University | Chonbuk National University | 767 | GRI | Korea Atomic Energy Research Institute | 247,795 |

**Table 1.** Top 10 agents for number of R&D Projects and R&D investment in Korean NIS

Table 1 lists the top innovation actors by number of R&D projects and funding. As can be seen, universities are mostly ranked first in terms of the number of R&D projects, indicating that universities conduct many innovation activities through a large number of R&D projects. However, Korea Aerospace Industries, Electronics and Telecommunications Research Institute (ETRI), and Korea Aerospace Research Institute (KARI), all classified as corporations and GRIs, are the top actors in terms of R&D expenditures. Seoul National University (SNU),

Yonsei University, Korea Advanced Institute of Science and Technology (KAIST), and Korea University also perform a large number of innovative activities in terms of research expenditures

## 4.2. Mapping the innovation capability of agents in Korean NIS

Using Eq. 1 and Eq. 2, we can measure the comparative advantage of actors in the NIS in S&T. In the NTIS data, 222 S&Ts are included in four major categories (Engineering, Natural Science, Human Science and Technology, Life Science). As an example, the RCA top 10 S&T for SNU and ETRI are summarized in Table 2.

| | SNU | | | ETRI | |
|---|---|---|---|---|---|
| S&T category | S&T | RCA | S&T category | S&T | RCA |
| Engineering | Ecosystem restoration/management | 6.83 | Engineering | Digital broadcasting | 13.80 |
| Natural Science | Model/Data analysis | 6.36 | Engineering | Broadband aggregation network | 12.93 |
| Human Science and Technology | Cognitive science | 5.64 | Engineering | Satellite/Radio | 13.87 |
| Life Science | Agricultural infrastructure engineering | 5.41 | Engineering | Telecommunications | 11.47 |
| Natural Science | Topology | 5.18 | Engineering | Power IT | 8.81 |
| Human Science and Technology | Brain neurobiology | 4.78 | Engineering | Telecom modules and components | 7.85 |
| Natural Science | Other math | 4.75 | Engineering | RFID/USN | 7.45 |
| Natural Science | Other geosciences | 4.67 | Engineering | Information protection | 7.24 |
| Natural Science | Algebra | 4.52 | Engineering | U-Computing | 6.94 |
| Natural Science | Geochemistry | 4.52 | Engineering | ITS/Telematics | 6.86 |

**Table 2.** Top 10 RCA for S&T for SNU and ETRI

As shown in Table 2, SNU has RCAs greater than 1 in various fields such as Engineering, Natural Science, Human Science and Technology, and Life Science, indicating innovation capability in these fields. In contrast, ETRI, a mission-oriented GRI for electronic communication technology, concentrates its capabilities in the field of Engineering, which is related to information and communication technology. The actor–activity relationships in the Korean NIS were constructed as an specialization matrix, $M$. By visualizing the constructed M matrix separately by UIG [1], it is possible to understand the distribution of innovation capabilities by UIG.

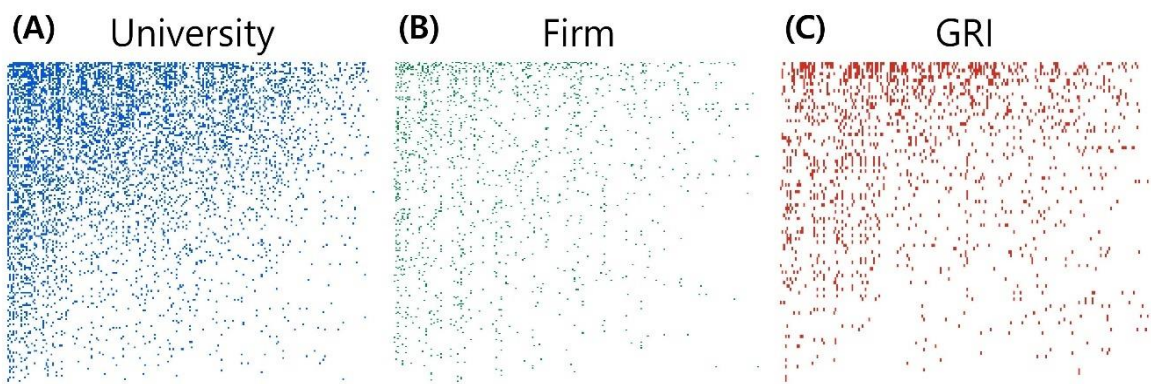

**Figure 2. Specialization matrix by innovation agent types**. (A) University, (B) Firm, and (C) Government Research Institute (GRI) institutions are visualized as binary matrices, where rows represent institutions and columns represent science and technology fields. Each blue (A), green (B), or red (C) dot indicates that the corresponding institution has a Revealed Comparative Advantage (RCA≥1) in that specific field. All matrices are ordered by decreasing

---

[1] Perform a visualization on the classifications for universities, firms, and GRI. There are 87 actors in other classifications, and 45 organizations in duplicate classifications

activity across both rows and columns. This visualization highlights structural differences in specialization and diversification patterns across the three agent types.

Figure 2A shows 181 universities in rows and 222 S&Ts in columns, sorted from top to bottom in order of universities with more innovation capabilities and from left to right in order of S&Ts where actors have more RCA. We depicted S&Ts where universities have innovation capability (RCA≥1) in red and those where they do not in white. Figure 2A shows a triangular shape with a large number of red points on the upper left and a large number of white points on the lower right. This suggests that, within the Korean NIS, universities with more innovation capabilities have capabilities in ubiquitous S&Ts in which many innovation actors have capabilities, as well as some non-ubiquitous S&Ts. However, universities that do not have many innovation capabilities have capabilities only in ubiquitous S&Ts.

Figure 2B shows the 269 firms in rows. Unlike universities, firms do not exhibit a triangular distribution. This means even firms with a large number of innovation capabilities (top of the matrix) do not have capabilities in S&Ts where other firms have a large number of capabilities (left side of the matrix), and even firms with a small number of innovation capabilities (bottom of the matrix) may have capabilities in S&Ts where few firms have capabilities (right side of the matrix). Hence, firms in the Korean NIS do not exhibit a uniform tendency in building innovation capabilities, but rather build innovation capabilities in differentiated S&Ts.

Figure 2C shows the visualization results for 95 GRIs. The distribution of GRIs is somewhere between the triangular distribution of universities and random distribution of companies. It can be seen that GRIs at the top of the matrix have more diverse capabilities, while those at the bottom of the matrix have differentiated capabilities. This can be explained by the fact that GRIs play two heterogeneous roles (Gulbrandsen, 2011), either conducting specialized research in a specific S&T or providing skills/knowledge to meet the needs of firms in various S&T.

## 4.3. Diversity and ubiquity in the Korean NIS

| Diversity (Descending Order) | | | Average Ubiquity (Ascending Order) | | |
|---|---|---|---|---|---|
| Category | Name | $k_{a,0}$ | Category | Name | $k_{a,1}$ |
| University | Seoul National University | 99 | Firm | Korea Aerospace Industries Co. | 7 |
| University | Korea University | 96 | Others | Korea Space Technology Promotion Association | 12 |
| University | Hanyang University | 89 | Others | Korea Atomic Energy Agency | 14.5 |
| University | Pusan National University | 84 | GRI | Korea Nuclear Cooperation Foundation | 15 |
| University | Kyungpook National University | 79 | Firm | Future and Challenge Co. | 15.6 |
| University | Yonsei University | 78 | GRI | Korea Astronomical Institute | 17.7 |
| University | Kangwon National University | 77 | GRI | Science and Technology Policy Institute | 18 |
| University | Kyung Hee University | 76 | GRI | Korea Aerospace Research Institute | 18.5 |
| University | Central University | 76 | Firm | Laon Nex Tep Co. | 19.3 |
| University/GRI | Korea Advanced Institute of Science and Technology | 75 | GRI | Korea Institute of Nuclear Fusion Energy | 21.4 |

**Table 3**. Top 10 agents with the highest diversity and lowest ubiquity levels

Actors' diversity ($k_{a,0}$) and ubiquity ($k_{a,1}$) are utilized to measure the level of diversity and ubiquity of innovation capabilities in the Korean NIS. Table 3 lists actors' diversity and ubiquity. Actors with high diversity are mostly universities, indicating that they have capabilities in a wide range of S&Ts. In addition, actors with low ubiquity levels imply that they have competencies in specialized S&Ts where other actors do not have competencies. In practice, low-ubiquitious actors are mission-oriented GRIs or firms established for research in areas that are difficult for universities and normal firms to compete in, such as space and nuclear energy.

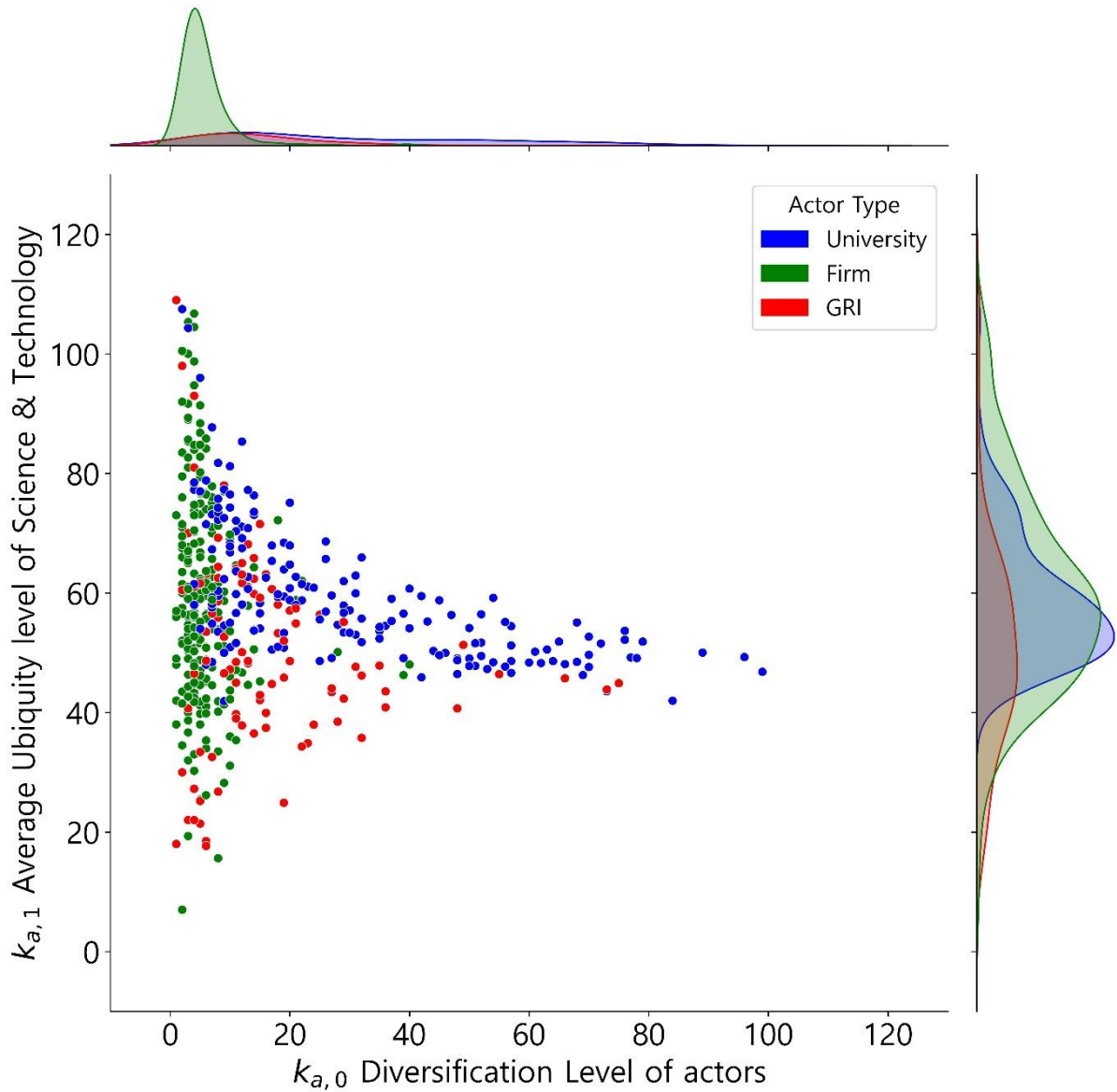

**Figure 3. Diversity - Average ubiquity level of innovation actors in Korean NIS.** The x-axis represents each actor's diversification level, while the y-axis shows the average ubiquity of the science and technology fields they are active in. Each dot represents an institution, and the color indicates the type of actor: university (blue), firm (green), or government research institute (GRI, red). A total of 72 other actor types is excluded from the plot for clarity.

Figure 3 is a a scatter plot with actor's diversity on the x-axis and ubiquity on the y-axis. The different colors represent the different types of actors: university, firm. As shown in Figure 2, the diversity and ubiquity of UIGs in Korea's NIS are distributed according to their respective characteristics. Universities show a large dispersion in terms of diversity, tending to have a higher average diversity than GRIs and firms. This suggests that universities are engaged in more diverse innovation activities on average within the NIS. Firms have a high dispersion of ubiquity at low diversity levels. This suggests that firms have competencies in a small number of S&Ts, and that the level of ubiquity of the S&Ts in which firms have competencies varies widely. Finally, GRIs show a medium level of diversity and ubiquity between universities and companies, reflecting the hybrid role of GRIs in the NIS.

The results in Figure 3 reflect the role of UIGs in NIS as suggested by previous literature. Universities play a role in generating new ideas and spawning new innovations in order to form the knowledge base of GRIs and firms (Datta et al., 2019; Kolomytseva and Pavlovska, 2020; Krishna, 2019). Therefore, within the NIS, universities perform significant innovation activities compared to GRIs and firms. This large extent of innovation activities is expressed through the high average level of diversity in Figure 3. Firms, by contrast, seek to deepen their technological knowledge in specific S&Ts or to innovate and thereby leverage the technological knowledge provided by GRIs and universities. Thus, firms have competencies in the specific S&Ts they utilize (Ahuja and Katila, 2001; Meyer et al., 1997; Ulrich, 1995). This is expressed by the low level of diversity in Figure 3. GRIs have both firm and university characteristics. GRIs are heterogeneous according to the degree of both characteristics (Gulbrandsen, 2011). GRIs play an intermediary role between universities in general knowledge production and firms in applied technology utilization. This intermediary role of GRIs is illustrated in Figure 3, where GRIs are distributed at the midpoint between universities and firms.

## 4.4. R&D Space of Korean NIS

In Eq. 5 and Eq. 6 of Section 3.3, we construct the R&D Actor Space by measuring the proximity between actors and forming links if the proximity level is 0.4 or higher. We first verify the appropriateness of vectorizing actors' innovation capabilities and measure the proximity between them by using cases from SNU and the ETRI.

| SNU | | | ETRI | | |
| --- | --- | --- | --- | --- | --- |
| Category | Name | Cosine Similairty | Category | Name | Cosine Similarity |
| University | Yonsei University | 0.583 | Firm/Others | Korea Electronics Technology Institute | 0.488 |
| University | Pusan National University | 0.548 | GRI | Korea Institute of Science and Technology Information | 0.399 |
| University | Chungnam National University | 0.544 | University | Korea University of Technology and Education | 0.366 |
| University | Sungkyunkwan University | 0.542 | University | Seoul National University of Science and Technology | 0.360 |
| University | Hanyang University | 0.540 | Firm | Autoa2z Co. | 0.354 |
| University | Korea University | 0.535 | Firm | SK Telecom Co. | 0.347 |
| University | Kyungpook National University | 0.513 | Firm | erangtek Co. | 0.333 |
| University | Chonbuk National University | 0.507 | Firm | Nepes Co. | 0.327 |
| University/GRI | Korea Advanced Institute of Science and Technology | 0.497 | Firm | Markany Co. | 0.323 |
| University | Dongguk University | 0.476 | Others | Gyeonggi-do Economic Science and Promotion Institute | 0.302 |

**Table 4**. Top 10 agents close to SNU and ETRI in terms of innovation capabilities

Table 4 lists the top 10 actors close to Seoul National University and ETRI in terms of innovation capabilities. Seoul National University has a high proximity (≥0.4) with many universities. In contrast, ETRI has low proximity (<0.4) with all but one innovation actor. This suggests that ETRI has a distinct innovation capability from other actors.

We construct the R&D Actor Space by measuring the proximity between all innovation actors, as defined in Table 4 and visualized in Figure 4. In this network, each node represents an innovation actor, with node size proportional to the actor's total R&D expenditure. Links between nodes indicate the degree of proximity, capturing the similarity in technological engagement among actors.

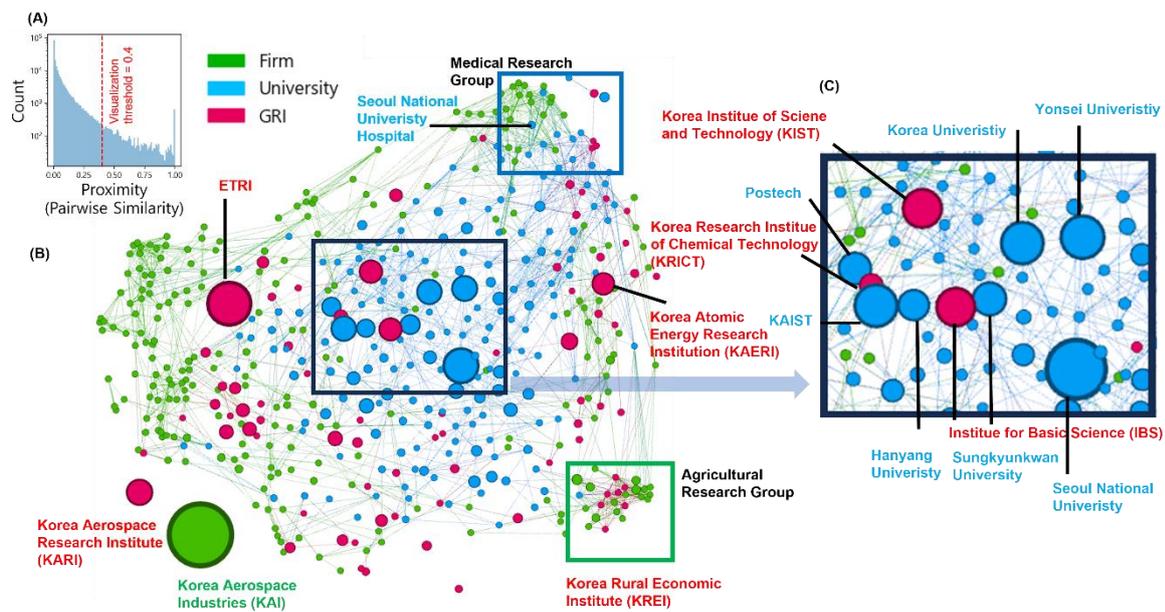

**Figure 4. R&D Actor Space of Korean NIS**. A network representation of 521 innovation actors based on pairwise similarity in their research specialization profiles (cosine similarity ≥ 0.4). Node size indicates R&D investment, and node color denotes actor type (green: firm; blue: university; red: GRI). Several capability-based clusters are visible, including medical, core academic, and peripheral specialized groups. See Table 4 and text for details.

Figure 4 presents this network view of the Korean NIS, encompassing 521 institutions. To enhance visualization clarity, actors categorized as "others" are excluded. Proximity between actors is calculated using cosine similarity of their research specialization profiles, and edges are formed when the similarity exceeds a threshold of 0.4. Node colors indicate actor type—green for firms, blue for universities, and red for government research institutes (GRIs). The histogram in the upper-left corner of the figure shows the distribution of pairwise similarity scores, with a red dotted line marking the threshold applied in the network construction. Several clusters of innovation actors with high capability proximity are visually identifiable. A distinct medical research cluster appears in the upper region, while a dense central group—labeled C—comprises major universities and GRIs, such as Seoul National University, KAIST, Yonsei University, and the Institute for Basic Science (IBS). In contrast, specialized actors such as ETRI, KARI, and KREI are located at the periphery, reflecting their differentiated innovation profiles.

This configuration reveals that the Korean NIS exhibits both concentrated interconnectivity among top academic institutions and structural differentiation among firms and GRIs. The R&D Actor Space thus provides a useful lens for understanding the relational distribution of innovation capabilities within the national system. R&D Actor Space intuitively shows that the Korean NIS is very densely connected among universities (blue) and somewhat sparsely connected among GRIs (red) and firms (green), indicating that the Korean NIS is highly closed among universities, and differentiated in terms of innovation capabilities among firms and GRIs representing industry and government.

Figure 5 illustrates the structural characteristics of the R&D Actor Space through degree-based network measures. In Figure 5A, the degree distribution $P(n)$ is plotted on log–log axes. The full distribution exhibits a curved pattern, as captured by the blue dashed quadratic fit, suggesting a deviation from a pure power-law form. To assess whether the upper tail of the distribution exhibits scale-free behavior, we fit a linear model to the subset of nodes with degree $n \geq 10$. The resulting estimate of the power-law exponent is $\gamma = 2.55$ with high statistical significance (p-value < 0.001), as shown by the red line. This indicates that while the network as a whole is not strictly scale-free, its high-degree regime approximates a power-law structure.

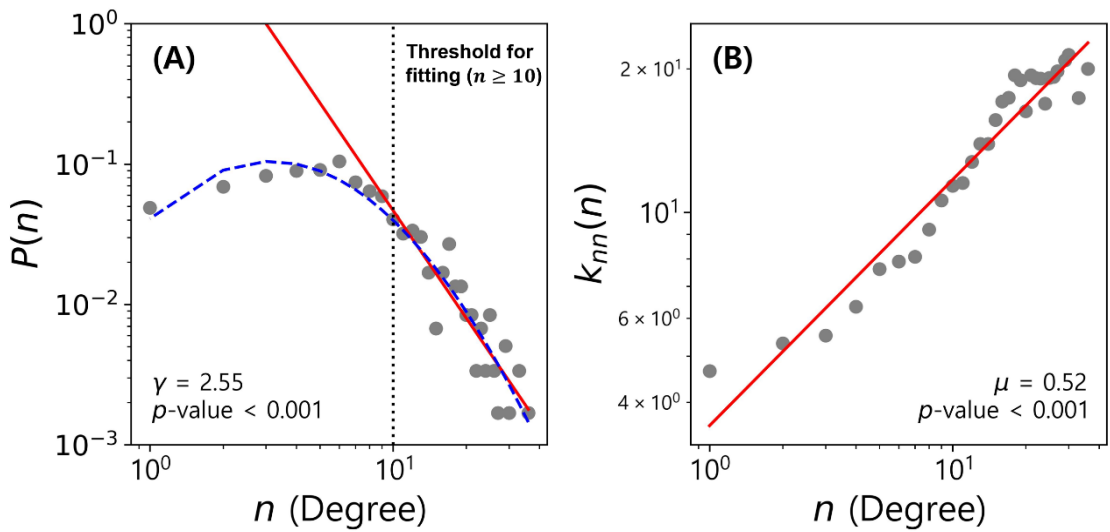

**Figure 5. Degree distribution and degree correlation in the R&D Actor Space.** (A) Degree distribution $P(n)$ and, (B) degree correlation $k_{nn}(n)$ plotted on log–log scales. Red lines indicate fitted linear regression lines, while the blue dashed curve in (A) represents a quadratic fit to the full distribution. Estimated slopes ($\gamma$ and $\mu$) and significance levels are annotated in each panel. See text for detailed interpretation.

Figure 5B presents the degree correlation, where the average degree of neighboring nodes $k_{nn}(n)$ is plotted against node degree $n$, both in log–log scale. The red line represents a linear fit, yielding an assortativity exponent of $\mu = 0.52$ (p-value < 0.001). This positive correlation indicates that high-degree actors tend to be connected with other high-degree actors, revealing an assortative structure in the R&D Actor Space. Together, the results suggest that while degree distribution is broadly heterogeneous, the network also exhibits structural clustering and inequality among highly connected actors.

This assortativity reflects a form of structural inequality within the innovation network—commonly referred to as "rich-club" behavior. To further investigate the origin of this inequality, we calculated assortativity ($\mu$) separately by actor type. we measure the degree correlation by the type of innovation actors and find the following results: firm ($\mu=0.618$), GRI ($\mu=0.507$), and university ($\mu=0.399$; p-value < 0.001). In other words, firm and GRI, which have a high degree correlation, are responsible for the inequality of innovation connectivity in the Korean NIS.

## 4.5. The hierarchical structure in Korean NIS

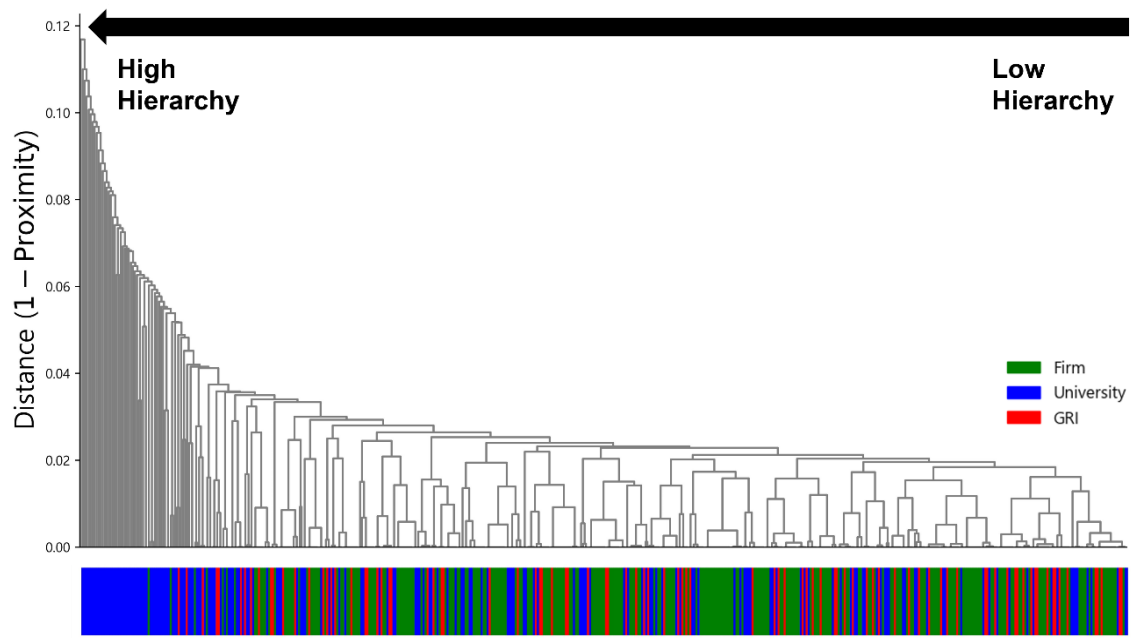

**Figure 7. Hierarchical structure of Korean R&D actor space.** The y-axis represents the proximity-based dissimilarity metric (1 − Proximity), and the x-axis shows individual institutions. The top black arrow illustrates the hierarchical structure of actors: actors on the left are positioned at higher levels of the hierarchy (i.e., more central or widely connected), while actors on the right are at lower levels. The colored bar at the bottom indicates each actor's category based on the TH model: green for firms, blue for universities, and red for government research institutes (GRIs).

To uncover the internal hierarchy among innovation actors in the Korean NIS, we apply hierarchical clustering based on actor-to-actor proximity in innovation capabilities. We adopt Ward's linkage algorithm, which is particularly effective in minimizing intra-cluster variance and producing interpretable dendrograms.

Figure 7 presents the resulting hierarchical clustering dendrogram. The vertical axis indicates proximity-based dissimilarity (1 − Proximity), and the horizontal axis displays an ordered list of institutions. The top arrow illustrates the overall gradient of hierarchical depth: actors positioned on the left remain unmerged until the later stages, indicating their placement at higher levels of the hierarchy (i.e., they are more distinct or central). In contrast, actors on the right are merged earlier, reflecting lower levels of hierarchical centrality.

We observe that universities tend to occupy the uppermost levels of the hierarchy. In particular, institutions such as Hanyang University, Seoul National University, Yonsei University, and Korea University are positioned at higher levels. These actors exhibit high diversity and connectivity across scientific domains, serving as central hubs in the actor space.

In contrast, GRIs and firms are more prevalent in the lower hierarchy. Specialized firms and GRIs such as the Korea Atomic Energy Research Institute (KAERI) and the National Institute for Mathematical Sciences (NIMS) tend to form narrow, late-merging branches, reflecting their highly focused and domain-specific innovation profiles. Firms often appear as dense bands along similar hierarchical levels, indicating the presence of multiple firm clusters with shared or similar innovation capabilities. These firm groups are relatively cohesive internally but maintain limited proximity to actors at higher hierarchical levels.

Taken together, the hierarchical clustering structure highlights the stratified nature of the Korean NIS, with central, highly diversified universities at the top and more specialized, peripheral actors such as GRIs and firms at the bottom. This structure reveals how different types of institutions contribute in distinct and layered ways to the overall innovation system.

## 5. Discussion and Conclusion

This study provides a novel methodological framework for analyzing the heterogeneous structure of the NIS by integrating the logic of ECT with relational network modeling. While the traditional TH model has offered a foundational view of university–industry–government (UIG) interactions, it has become increasingly inadequate for capturing the internal variation and stratification among innovation actors. Our findings demonstrate that the proposed network-based approach—centered on the construction of the R&D Actor Space—offers a meaningful alternative for understanding the diversity, connectivity, and hierarchy embedded within the NIS.

Empirically, we apply this approach to the Korean NIS using national R&D project data, validating the method's appropriateness across multiple dimensions. First, by measuring the RCA of institutions across S&T fields, we derive the specialization profiles of over 500 innovation actors. This enables us to construct a proximity-based network of institutions that we term the R&D Actor Space. This network captures relational similarity in innovation capabilities without assuming predefined actor categories, allowing for more granular structural analysis.

The results reveal several important characteristics of the Korean innovation system. Universities tend to exhibit both high diversity and low ubiquity in their innovation activities, positioning them as central hubs in the actor space. GRIs and firms, in contrast, are more heterogeneous: some act as peripheral specialists, while others form dense clusters of similarly positioned actors. The structural analysis of the R&D Actor Space confirms this stratification.

Degree-based metrics show that the network is assortative and exhibits a rich-club pattern among highly connected actors—particularly among universities and large GRIs.

The hierarchical clustering results further support these observations. By applying Ward's method to the actor-to-actor proximity matrix, we visualize a nested dendrogram that uncovers the relative depth of actor similarity. Universities such as Seoul National University, Yonsei University, Korea University, and Hanyang University occupy the highest levels of the hierarchy, reflecting their broad engagement across multiple S&T domains. In contrast, specialized GRIs (e.g., KAERI, NIMS) and firms tend to occupy the lower layers, either forming narrow branches or concentrated bands along specific domains.

From a methodological standpoint, the combination of RCA-based specialization metrics and proximity-based clustering demonstrates strong analytical power in representing the relational heterogeneity of the NIS. Unlike categorical typologies that obscure internal variation, our framework preserves fine-grained differences and reveals structural properties that are otherwise difficult to detect. Also, this study provides an alternative methodology for ECT's application on finding the pathways of innovation capability development.

This approach holds important implications for future research. First, the R&D Actor Space can serve as a foundational tool for monitoring shifts in national innovation strategies, institutional reconfiguration, or policy interventions over time. Second, the hierarchical and modular patterns observed here invite further investigation into the mechanisms that drive actor positioning, such as funding allocation, collaboration intensity, or mission specificity. Lastly, the generalizability of the method allows for comparative analysis across countries, enabling cross-national benchmarking of innovation system architectures.

In conclusion, this study offers a scalable and interpretable framework for observing the internal dynamics of innovation systems. By embracing relational heterogeneity and moving beyond static classifications, it provides both theoretical and empirical tools to advance our understanding of how innovation actors are organized—and how they evolve—within complex national systems.

**Acknowledgement**

This work was supported by the Korea Institute of S&T Evaluation and Planning (KISTEP) grant funded by the Korea government (MSIT) (No. AR23030).

**Conflicts of interest**

The authors declare that they have no conflict of interest.

**Funding**

This work was supported by the Korea Institute of S&T Evaluation and Planning (KISTEP) grant funded by the Korea government (MSIT) (No. AR23030).

**Author contributions**

**Dawoon Jeong:** Conceptualization, Data curation, Formal analysis, Funding acquisition, Investigation, Methodology, Software, Supervision, Validation, Visualization, Writing - original draft, Writing - review & editing

**Taewon Kang:** Supervision, Validation, Writing - review & editing

**Saerom Si:** Conceptualization

**Sangnam Lee:** Conceptualization, Funding acquisition, Project administration, Resources

**Wonsub Eum:** Supervision, Investigation, Writing - review & editing